\documentclass[12pt,preprint]{aastex}
\def\degpoint{\ifmmode ^{\rm{o}}\!. \else $^{\rm{o}}\!.$\fi}

\newcommand{\ms}{\mbox{m\,s$^{-1}$}}

\newcommand{\Msun}{\mbox{M$_{\odot}$}}

\newcommand{\Mjup}{\mbox{M$_{\rm Jup}$}}

\newcommand{\gtsimeq}{\raisebox{-0.6ex}{$\,\stackrel
         {\raisebox{-.2ex}{$\textstyle >$}}{\sim}\,$}}

\begin{document}

\title{Resonances Required: Dynamical Analysis of the 24\,Sex and 
HD\,200964 Planetary Systems }

\author{Robert A.~Wittenmyer\altaffilmark{1}, Jonathan 
Horner\altaffilmark{1}, C.G. Tinney\altaffilmark{1} }
\altaffiltext{1}{Department of Astrophysics, School of Physics, Faculty 
of Science, The University of New South Wales, 2052, Australia}

\email{
rob@phys.unsw.edu.au}

\shortauthors{Wittenmyer, Horner, \& Tinney}

\begin{abstract}

\noindent We perform several suites of highly detailed dynamical simulations 
to investigate the architectures of the 24\,Sextantis and HD\,200964 
planetary systems.  The best fit orbital solution for the two planets in 
the 24\,Sex system places them on orbits with periods that lie very 
close to 2:1 commensurability, while that for the HD\,200964 system 
places the two planets therein in orbits whose periods lie close to a 
4:3 commensurability.  In both cases, the proposed best-fit orbits are 
mutually crossing - a scenario that is only dynamically feasible if the 
planets are protected from close encounters by the effects of mutual 
mean motion resonance.  Our simulations reveal that the best fit orbits 
for both systems lie within narrow islands of dynamical stability, and 
are surrounded by much larger regions of extreme instability.  As such, 
we show that the planets are only feasible if they are currently trapped 
in mutual mean-motion resonance - the 2:1 resonance in the case of 
24\,Sex~b and c, and the 4:3 resonance in the case of HD\,200964~b and 
c.  In both cases, the region of stability is strongest and most 
pronounced when the planetary orbits are mutually coplanar.  As the 
inclination of planet c with respect to planet b is increased, the 
stability of both systems rapidly collapses.
\end{abstract}


\section{Introduction}

Systems of multiple extrasolar planets provide rich laboratories for 
testing theories of planet formation and dynamical evolution.  
Radial-velocity planet-search programs, with observational baselines 
approaching 20 years, often discover that apparent single-planet systems 
show evidence for additional planets.  These discoveries are due to a 
combination of increased time coverage and ongoing improvements in 
measurement precision arising from innovation and refinement in 
instrumentation, observing strategies, and analysis techniques 
\citep{vogt10,mayor11}.  The former enables the detection of 
Jupiter-analog planets with orbital periods $P\gtsimeq$10 yr 
\citep{wright09,jones10,142paper}, and the latter permits the robust 
detection of planets of ever lower mass.  \citet{wright09} showed that 
at least 28\% of planetary systems surveyed by Doppler programs host 
multiple planets.  Recent results from the \textit{Kepler} spacecraft, 
with some 885 multiply-transiting planet candidates \citep{fabrycky12}, 
have vastly increased the pool of these fascinating and valuable 
planetary systems.

A number of recent studies have highlighted the need for observational 
detections of multiple-planet systems to be supported by dynamical 
investigations that test whether the orbits of the proposed planets are 
dynamically feasible.  These investigations can reveal that the 
planetary system configuration as reported is catastrophically unstable 
\citep{horner11, HUAqr, hinse12a, g12, HWVir}.  Alternately, detailed 
dynamical simulations of systems close to mutual mean-motion resonance 
(MMR) can provide important additional constraints on the parameters of 
the planets (e.g. \citet{NNSer,155358paper}).  This is critical because 
any commensurability between the orbital periods of two planets can 
result in either extreme stability or instability, depending on the 
precise orbits of the planets involved.  For example, 
\citet{155358paper} presented revised orbits for the two Jovian planets 
in the HD\,155358 system \citep{cochran07}, potentially placing them in 
mutual 2:1 MMR.  The dynamical stability analysis of \citet{155358paper} 
showed that, for the orbital architecture presented in that work, the 
2:1 MMR was indeed a region of long-term stability, serving as further 
evidence that the new orbital parameters are correct.  In the HD\,204313 
three-planet system \citep{seg10}, a similar dynamical analysis reveals 
that the observed two outermost Jovian planets must be trapped in mutual 
3:2 MMR \citep{204313}, an unusual architecture which dynamical mapping 
showed to be an island of extreme stability in the system.  In that 
case, the dynamical analysis was critically necessary to constrain the 
HD\,204313 system architecture, demonstrating that the two gas giant 
planets are locked in the 3:2 MMR.

The intriguing results of these detailed stability investigations have 
prompted us to take a close look at other planetary systems which appear 
to be in or near low-order MMRs.  In recent years, a number of surveys 
\citep{hatzes05, sato05, johnson06b, doellinger07, johnson11b, 
47205paper} have discovered a significant number of planetary systems 
orbiting intermediate-mass stars ($M_{*}>1.5$\Msun).  These stars have 
proven to be a fertile hunting ground for interesting planetary systems.  
Of particular interest are the two low-order resonant giant planet pairs 
in the 24~Sex (2:1) and HD\,200964 (4:3) systems \citep{johnson11}.  For 
both systems, the best-fit radial-velocity solutions result in crossing 
orbits, an architecture which can only be dynamically stable on long 
timescales if the planets are protected from mutual close encounters by 
resonant motion.

In their study of 24\,Sex and HD\,200964, \citet{johnson11} performed 
small-scale \textit{n}-body dynamical simulations which showed, as 
expected, that the long-term stable solutions were restricted to the 2:1 
(24\,Sex) and 4:3 (HD\,200964) mean-motion resonances.  The authors 
noted that their simulation results could not yet conclusively confirm 
that the planets were in resonance, and urged further, more detailed 
dynamical investigation.

Resonant protection, such as that proposed for the 24~Sex and HD\,200964 
systems, is well known and studied in our Solar system.  A wealth of 
Solar system objects move on orbits that would be highly unstable, were 
it not for the influence of such resonances, including the Hilda 
asteroids \citep{HILDA1}, Jovian and Neptunian Trojans \citep{Troj1, 
Troj2, Troj3, Troj4} and, most famously, the dwarf planet Pluto and its 
brethren, the Plutinos\citep{Plut1, Plut2}.  Taken in concert with the 
growing catalogue of resonant exoplanets, these populations highlight 
the important role resonant dynamics plays in the formation and 
evolution of planetary systems.  Indeed, astronomers studying the 
formation and evolution of our Solar system have learned a great deal 
about the extent, pace, and nature of the migration of the giant planets 
through studies of the system's resonant small body populations (e.g. 
\citet{Plut1, Troj1, NTmig, HILDAMIG}), revealing that the formation of 
our Solar system was most likely a relatively chaotic process.

In this paper, we apply our highly detailed dynamical analysis 
techniques to the extremely interesting 24\,Sex and HD\,200964 planetary 
systems.  In Section 2, we briefly describe the simulation methods and 
initial conditions.  Section 3 gives the results, and we give our 
conclusions in Section 4.

\section{Numerical Methods}

To study the dynamics of the two planetary systems proposed in 
\citet{johnson11}, we performed two main suites of dynamical simulations - 
one for each of the planetary systems studied.  We followed the strategy 
we have successfully employed to study the dynamics of a number of other 
exoplanetary systems (e.g. \citet{HR8799, horner11, HUAqr, NNSer}), and 
followed the dynamical evolution of a large number of different 
architectures for each system using the Hybrid integrator within the 
\textit{n}-body dynamics package MERCURY \citep{Mercury}.  In each case, 
we placed the better constrained of the two planets in question 
(24~Sex~b and HD~200964~b) on its nominal best-fit orbit at the start of 
our integrations.  The orbital parameters of the planets simulated in 
this work are shown in Table~\ref{planetparams} \footnote{The orbital 
solutions for the planets were taken from the Exoplanets Data Explorer 
website, http://exoplanets.org, on the 27th March 2012. We note that 
there are very small differences between the elements presented in that 
explorer and those presented in \citet{johnson11}. These are, however, 
far smaller than the 1$\sigma$ errors on the values, and have no effect 
on our results or conclusions}.  For the other planet (24~Sex~c and 
HD~200964~c), we then tested 41 unique orbital semi-major axes, 
distributed uniformly across the full $\pm 3 \sigma$ range allowed by 
the uncertainties in that planet's orbit.  For each of these 41 possible 
semi-major axes, we tested 41 unique eccentricities, again spread evenly 
across the full $\pm 3\sigma$ range of allowed values.  For each of 
these 1681 $a-e$ values, we tested 15 values of the longitude of the 
planet's periastron, $\omega$, and 5 values of its mean anomaly, $M$, 
each spanning the appropriate $\pm 3 \sigma$ error ranges.  In this way, 
a total of 126,075 potential architectures were tested for each system.

\begin{figure}
\plotone{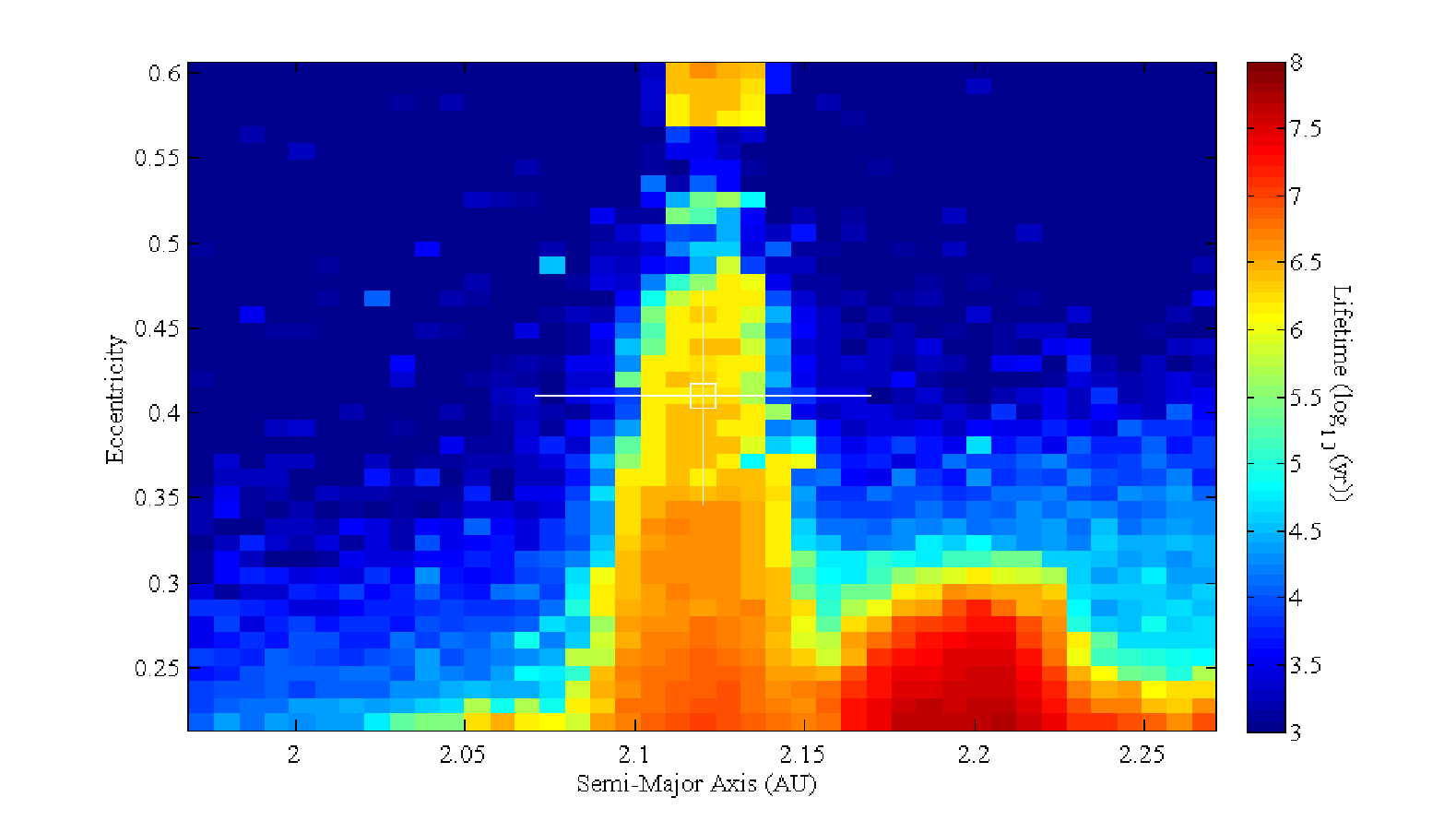}
\caption{Dynamical stability for the 24 Sex system as a function of the 
initial semimajor axis and eccentricity of the outer planet.  The 
nominal best-fit orbit for that planet is marked by the open square, and 
the 1$\sigma$ uncertainties are shown by the crosshairs.  The 2:1 
resonance appears as a narrow strip of stability which coincides with 
the outer planet's best-fit orbit \citep{johnson11}.}
\label{24sex}
\end{figure}

In each of our simulations, we followed the evolution of the two planets 
involved for a period of up to 100 Myr, until they were either ejected 
from their system, collided with one another, or were thrown into their 
central star.  The times at which collisions and ejections occurred were 
recorded, which allowed us to create dynamical maps of the system's 
stability (Figures~\ref{24sex} and \ref{200964}).

In addition to the main suites of integrations discussed above, we 
performed five additional suites of integrations for each of the two 
planetary systems to investigate the influence that the mutual 
inclinations of the planetary orbits would have on their stability.  In 
this, we followed \citet{HUAqr}, and considered cases where planet c was 
initially moving on an orbit inclined by 5, 15, 45, 135 and 180 degrees 
with respect to that of planet b.  Due to the significant computational 
overhead in performing such runs, the resolution of these subsidiary 
investigations was lower than for the main runs.  For each of these 
additional runs, a total of 11,025 trials were carried out, yielding the 
results shown in Figures 2 and 4.

\section{Results and Discussion}

\subsection{The 24 Sextantis System}

Figure~\ref{24sex} shows the results of our dynamical simulations for 
the 24\,Sex 2:1 resonant system.  At each point in the $(a,e)$ grid, the 
small colored cell represents the mean survival time of 75 variants of 
the two-planet system -- each with a unique initial combination of 
longitude of periastron and mean anomaly for the outer planet.  The 
initial orbital parameters for the inner planet were held fixed, as 
noted in the previous section.  The best-fit orbit for the outer planet 
is shown as an open box with crosshairs indicating the 1$\sigma$ 
uncertainties in semi-major axis and eccentricity.  The orbital solution 
of \citet{johnson11} lies directly on the narrow region of stable 
orbits, with mean survival times $>10^6$ yr.  This is strong evidence 
that the planets are truly in 2:1 resonance -- that the resonance is 
\textit{required} for the stability of the system.  Nearly all of the 
surrounding parameter space is highly unstable, with survival times 
typically less than $10^4$ yr.  An additional small region of stability 
can be seen centered at $a\sim2.2$ AU, for eccentricities below around 
0.3.  That region is the result of an overlapping web of weak, 
high-order resonances that congregate in the region 2.174 - 2.243 
AU\footnote{More explicitly, the following commensurabilities are found 
in that region - 31:15 at 2.174 AU; 29:14 at 2.177 AU; 27:13 at 2.181 
AU; 25:12 at 2.186 AU; 23:11 at 2.191 AU; 21:10 at 2.197 AU; 19:9 at 
2.205 AU; 17:8 at 2.215 AU; 15:7 at 2.227 AU; 28:13 at 2.235 AU; 13:6 at 
2.244 AU. At low eccentricities, such resonances can help to ensure the 
stability of orbits that would otherwise be somewhat unstable, as has 
been observed for both Solar system objects and exoplanetary systems 
(e.g. \citet{SUBTROJ, 155358paper, Kepler36}).}.  Interestingly, exactly 
the same feature can be seen in Figure 9 of \citet{155358paper}.  The 
HD~155358 planetary system is somewhat analagous to that around 24~Sex, 
in that it features two planets that are most likely trapped in mutual 
2:1 MMR.  In that case, HD\,155358c is most likely moving on an orbit 
that is somewhat less eccentric than that proposed for 24~Sex~c, and so 
their Figure~9 shows the dynamical stability of that system to lower 
eccentricities.  As the eccentricity of the outermost planet falls, the 
broad region of stability offered by these overlapped high-order 
resonances broadens until it merges with that offered by the protection 
of the 2:1 MMR.  We note, however, that the stability region for 
24~Sex~c offered by those higher order resonances lies well away from 
the central $\pm 1 \sigma$ of the allowed orbital architectures for the 
system.  Hence it seems far more reasonable to conclude that the planets 
in this system are, most likely, trapped in mutual 2:1 MMR.

\begin{figure}
\plotone{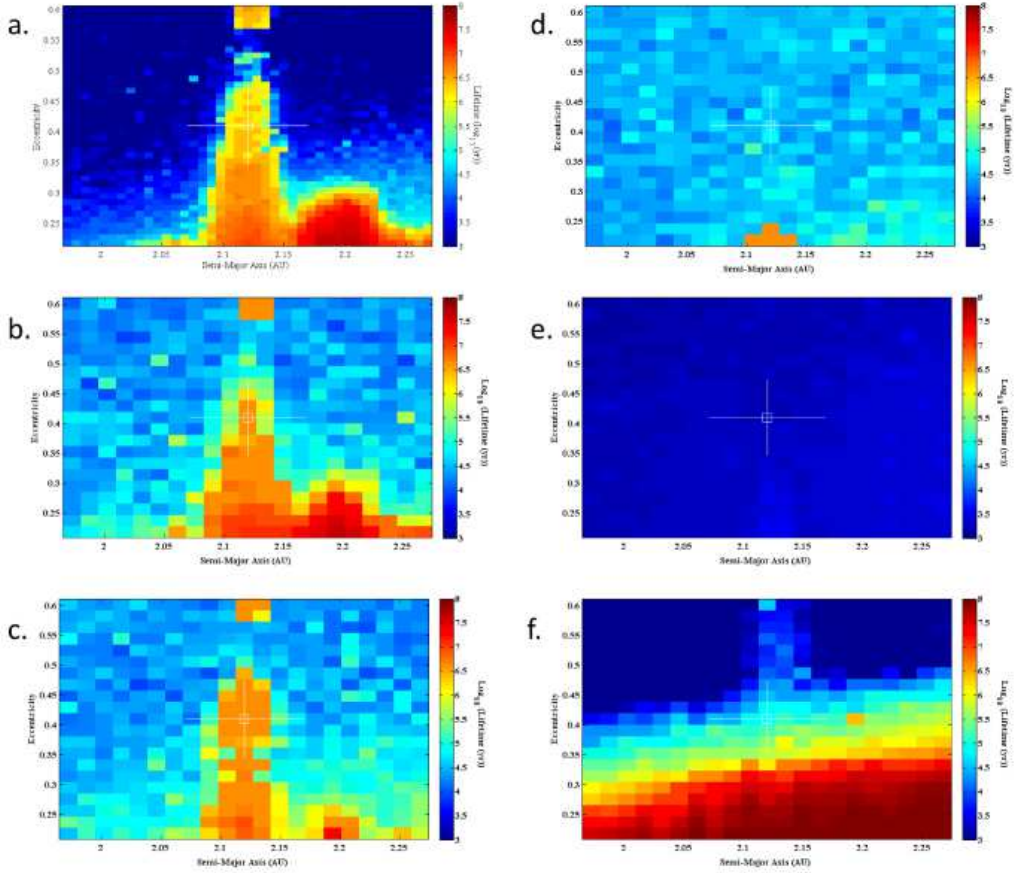}
\caption{Dynamical stability for the 24~Sex system, but for six values 
of inclination between the two planets. Panels (a) through (f) represent 
mutual inclinations of 0, 5, 15, 45, 135, and 180 degrees, respectively.  
Panel (a) is a duplicate of Figure 1, shown here for ease of comparison.  
As in previous figures, the color bar represents the log of the mean 
survival time. }
\label{24sexinclin}
\end{figure}

If the two planets were scattered to their present locations by a 
distant body, or by mutual chaotic interactions during their migration 
\citep{barnes11, mw02}, it is possible that they have some non-zero 
inclination relative to each other.  To explore the effect of 
mutually-inclined scenarios, we performed a subsidiary suite of 
integrations for a range of mutual inclination angles: 5, 15, 45, 135, 
and 180 degrees (i.e.~coplanar but retrograde).  These runs were set up 
as previously described, except at lower resolution: each scenario 
consists of a grid of 21 values of $a$, 21 of $e$, 5 of $\omega$, and 5 
of $M$ (11025 total trial systems).  The results are given in 
Figure~\ref{24sexinclin}, and show that the system becomes generally 
more unstable when the planets depart from a prograde, coplanar 
configuration.  Notably, the two retrograde scenarios had dramatically 
shorter lifetimes (panels e and f).  We note that while the retrograde 
coplanar case (panel f) shows a long-term stable region in the lower 
right, that region is quite far from the $1\sigma$ uncertainty on the 
orbit, much like our previous work on the proposed HU Aquarii planetary 
system \citep{HUAqr}.

\subsection{The HD\,200964 System}

\begin{figure}
\plotone{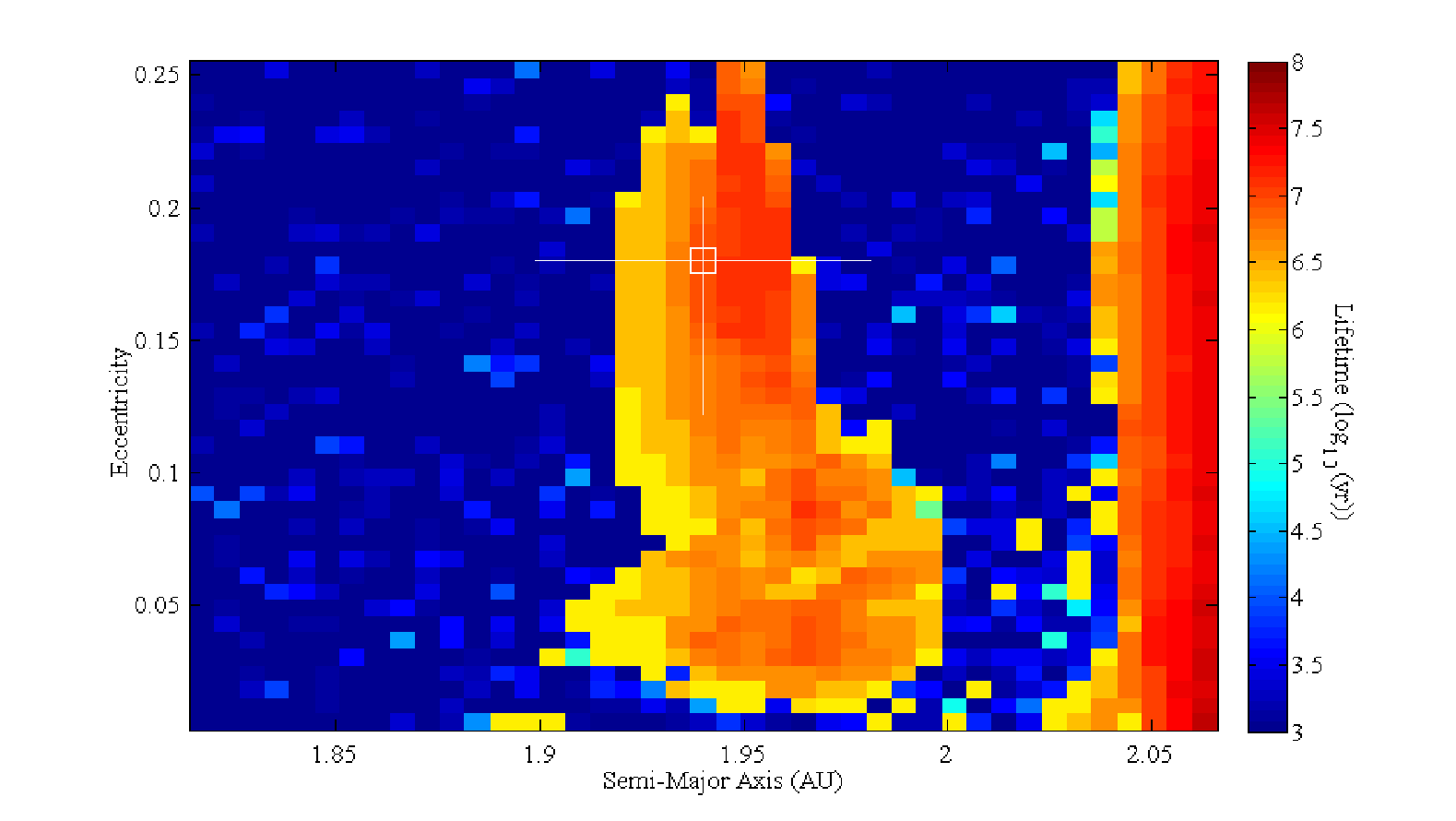}
\caption{Dynamical stability for the HD\,200964 system as a function of the
initial semimajor axis and eccentricity of the outer planet.  The
nominal best-fit orbit for that planet is marked by the open square, and
the 1$\sigma$ uncertainties are shown by the crosshairs.  The 4:3
resonance appears as a narrow strip of stability which coincides with
the outer planet's best-fit orbit \citep{johnson11}.}      
\label{200964}
\end{figure}

Figure~\ref{200964} shows the results for the HD\,200964 4:3 resonant 
system.  As in Figure~\ref{24sex}, the open box with crosshairs shows 
the best-fit parameters for the outer planet \citep{johnson11}.  Once 
again, we see that the radial-velocity solution for this system, with 
the planets in a 4:3 resonance, lies within a narrow region of orbital 
stability surrounded by highly unstable parameter space.  As was the 
case for the 24~Sex system, this dynamical map shows that the resonance 
is required for the system's stability.  An interesting feature of the 
4:3 resonant protection is that the region of stability does not extend 
all the way to zero eccentricity - in other words, some small, non-zero 
eccentricity is required for HD\,200964c to be dynamically stable within 
the 4:3 MMR with HD\,200964b.  This instability at very low 
eccentricities is observed in the Solar system's Plutino population 
(trapped in 3:2 MMR with Neptune; see e.g. Fig. 6 of \citet{204313}), 
and was also observed in some of the more extreme integrations of the 
planetary system orbiting HD~142 \citep{142paper}.


At the far right-hand edge of the allowed range, for $a>2.05$\,AU, we 
see a highly stable region at all tested eccentricities.  This feature 
is a common outcome of dynamical stability results at moderate and low 
eccentricities, located just interior to the location of the 2:1 MMR.  A 
similar feature can be seen in Fig~1 of \citet{horner11}, for the 
otherwise dynamically unfeasible HU~Aquarii planetary system.  It 
typically represents the inner edge of the region for which dynamically 
stable solutions become the norm, rather than the exception, apart from 
resonant interactions.  At greater separations, this region of stability 
extends to ever greater eccentricities, since the boundary between 
stable and unstable solutions is determined, for the non-resonant case, 
by the closest approach distance between the two planets in question.  
In \citet{horner11}, this inner edge was discussed in terms of the Hill 
Radius, $R_H$, of the more massive, innermost planet proposed in that 
work.  The divide between stable and unstable orbits was found to follow 
a line that (roughly) followed a line of constant periastron distance 
for the outermost planet, centred on a periastron distance between 3 and 
5 Hill radii beyond the orbit of the innermost planet.  For objects on 
near-circular orbits, the Hill Radius can be approximated as

$R_H = a \sqrt[3]{m \over {3M}}$

where $a$ and $m$ are the semi-major axis and mass of the planet in 
question, and M is the mass of the central star.  For HD~200964, the 
sharp boundary between unstable and stable orbits at around 2.05 AU once 
again lies between three and four Hill radii beyond the orbit of the 
innermost planet.  Interestingly, we note that \citet{MutHil} find that 
a system of two planets on low eccentricity, low inclination orbits ``is 
stable with respect to close encounters if the initial semi-major axis 
difference, $\Delta$, measured in mutual Hill radii, $R_H$, exceeds 
$2\sqrt{3}$, due to conservation of energy and angular momentum.'' The 
mutual Hill radius, $R_{HM}$ is defined as

$R_{HM} = \sqrt[3]{({{m_b + m_c} \over {3 M}})}({{a_b + a_c} \over 2} )$

where $m_b$, $m_c$, $a_b$, and $m_c$ are the masses and semi-major axes 
of planets b and c, and $M$ is the mass of the central star.  Given the 
criterion that the inner edge of the stable region should be found when 
the planets are separated by a distance of $2\sqrt{3}$ times their 
mutual Hill radius, it is therefore trivial to work out where the inner 
edge of this stable region should be expected to lie, for the case where 
the orbits are circular.  Holding the location of planet b fixed, we 
thus find that the inner edge of the stable region should lie at $a = 
2.15$AU, a little more distant than that observed in our 
Figure~\ref{200964}.  Once again, however, it should be noted that this 
stable region lies well beyond the central $\pm 1 \sigma$ region, and so 
represents a significantly less likely architecture for the HD\,200964 
system.

As for the 24~Sex system, we also considered mutually-inclined 
scenarios, running a further series of simulations as described in 
Section 3.1.  The results for the HD\,200964 system are the same: the 
mean lifetime decreases significantly when the planets are inclined by 
more than 90 degrees with respect to each other (retrograde orbits).  
Again, the most stable configuration was prograde and coplanar (panel a, 
identical to Figure~3).

\section{Conclusions}

Both the 24\,Sex and HD\,200964 systems host giant planets in close 
resonances \citep{johnson11}.  We have performed detailed dynamical 
simulations testing the 3$\sigma$ range of allowed parameter space for 
these two systems.  Our results have further constrained the orbital 
parameters, with the best-fit solutions falling directly in narrow 
($\sim 1\sigma$ width) strips of long-term stability.  We also find that 
the stability of both systems is strongly dependent on the mutual 
inclinations of the planets involved, with coplanar orbits offering by 
far the greatest potential for dynamically stable solutions to be found.  
This work demonstrates the utility of such dynamical mapping for better 
understanding the architectures of multiple-planet systems.  The results 
of this work confirm that the resonant configurations are indeed 
required for long-term stability in the 24\,Sex and HD\,200964 systems. 
This adds them to a very short list of low-order-resonant exoplanetary 
systems, which are extremely valuable test cases for understanding 
giant-planet formation and migration processes.

\acknowledgements

We gratefully acknowledge fruitful discussions with John Johnson which 
helped in the development of this paper.  RW is supported by a UNSW 
Vice-Chancellor's Fellowship.  JH is supported by ARC Grant DP0774000.  
The work was supported by iVEC through the use of advanced computing 
resources located at the Murdoch University, in Western Australia.  This 
research has made use of NASA's Astrophysics Data System (ADS), and the 
SIMBAD database, operated at CDS, Strasbourg, France.  This research has 
also made use of the Exoplanet Orbit Database and the Exoplanet Data 
Explorer at exoplanets.org.


\begin{deluxetable}{llllllll}
\tabletypesize{\scriptsize}
\tablecolumns{8}
\tablewidth{0pt}
\tablecaption{Planetary System Parameters }
\tablehead{
\colhead{Planet} & \colhead{Period} & \colhead{$T_0$} & \colhead{$e$} & 
\colhead{$\omega$} & \colhead{K } & \colhead{M sin $i$ } & \colhead{$a$ } \\
\colhead{} & \colhead{(days)} & \colhead{(JD-2400000)} & \colhead{} & 
\colhead{(degrees)} & \colhead{(\ms)} & \colhead{(\Mjup)} & \colhead{(AU)}
}
\startdata
\label{planetparams}   
24 Sex b & 455.2$\pm$3.2 & 54758$\pm$30 & 0.184$\pm$0.029 & 227$\pm$20 & 
33.2$\pm$1.6 & 1.6$\pm$0.2 & 1.338$\pm$0.024 \\
24 Sex c & 910$\pm$21 & 54941$\pm$30 & 0.412$\pm$0.064 & 172$\pm$9 & 
23.5$\pm$2.9 & 1.4$\pm$0.2 & 2.123$\pm$0.049 \\
HD 200964 b & 613.8$^{+1.3}_{-1.4}$ & 54916$\pm$30 & 
0.040$^{+0.04}_{-0.02}$ & 288$^{+47}_{-112}$ & 34.5$^{+2.7}_{-1.5}$ & 
1.84$^{+0.14}_{-0.08}$ & 1.601$\pm$0.002 \\
HD 200964 c & 825.0$^{+5.1}_{-3.1}$ & 55029$\pm$130 & 
0.181$^{+0.024}_{-0.058}$ & 182.6$^{+67.7}_{-57.1}$ & 15.4$\pm$3.2 & 
0.895$^{+0.123}_{-0.063}$ & 1.944$\pm$0.041 \\
\enddata 
\end{deluxetable}

\begin{figure}
\plotone{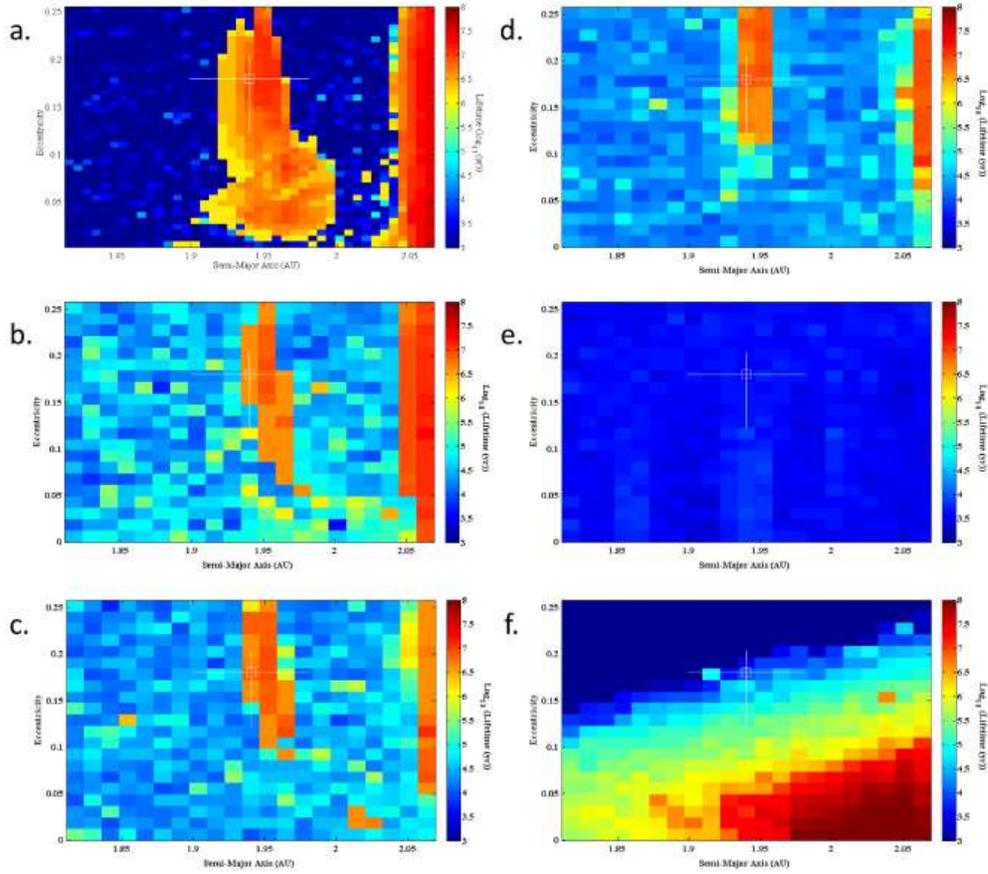}
\caption{Same as Figure~\ref{24sexinclin}, but for the HD\,200964 
system. As for 24~Sex, mutual-inclination scenarios are much less stable 
than the prograde-coplanar scenario in panel (a). }
\label{200964inclin}
\end{figure}

\end{document}